\documentclass[conference]{IEEEtran}
\IEEEoverridecommandlockouts
\usepackage{cite}
\usepackage{amsmath,amssymb,amsfonts}
\usepackage{algorithmic}
\usepackage{graphicx}
\usepackage{textcomp}
\usepackage{xcolor}
\def\BibTeX{{\rm B\kern-.05em{\sc i\kern-.025em b}\kern-.08em
    T\kern-.1667em\lower.7ex\hbox{E}\kern-.125emX}}

\usepackage{color}
\usepackage[hidelinks]{hyperref}
\usepackage{graphicx}
\usepackage{amsmath}
\usepackage{amsfonts}
\usepackage{mathrsfs}
\usepackage{xcolor}
\usepackage{tcolorbox}
\usepackage{mwe}
\usepackage{enumitem}
\usepackage{pifont}
\usepackage{tabularx}

\newcommand{\T}{^{\mbox{\tiny T}}}
\usepackage{url}

\newtheorem{definition}{\bf Definition}
\newtheorem{lemma}{\bf Lemma}
\newtheorem{assumption}{\bf Assumption}
\newtheorem{problem}{\bf Problem}
\newtheorem{theorem}{\bf Theorem}

\DeclareMathOperator{\Ker}{Ker}
\DeclareMathOperator{\Span}{span}

\newcommand{\R}{\mathbb{R}}

\let\leq\leqslant
\let\geq\geqslant

\newenvironment{proof}[1][Proof]%
{\par\noindent\textit{#1:\ }}%
{\hspace*{\fill} \rule{6pt}{6pt}}
\newenvironment{proof*}[1][Proof]%
{\par\noindent\textit{#1:\ }}{}

\IfFileExists{wasysym.sty}%
{
	
	\usepackage{wasysym}%
	}%
{}
\newenvironment{system}[1]%
{\setlength{\arraycolsep}{0.5mm}\left\{ \; \begin{array}{#1}}%
	{\end{array} \right.}
\newenvironment{system*}[1]%
{\setlength{\arraycolsep}{0.5mm} \begin{array}{#1}}%
	{\end{array}}

\begin{document}

\title{Weak state synchronization of homogeneous multi-agent systems with adaptive protocols\\
}

\author{\IEEEauthorblockN{1\textsuperscript{st} Anton A. Stoorvogel}
\IEEEauthorblockA{\textit{Department of Electrical Engineering} \\\textit{Mathematics
		and Computer Science}\\
\textit{University of Twente}\\
Enschede, The Netherlands\\
A.A.Stoorvogel@utwente.nl}
\and
\IEEEauthorblockN{2\textsuperscript{nd} Ali Saberi}
\IEEEauthorblockA{\textit{School of Electrical Engineering and Computer
		Science} \\
\textit{Washington State University}\\
Pullman WA, USA\\
saberi@wsu.edu}
\and
\IEEEauthorblockN{3\textsuperscript{rd} Zhenwei Liu}
\IEEEauthorblockA{\textit{College of Information Science and Engineering} \\
\textit{Northeastern University}\\
Shenyang, China \\
liuzhenwei@ise.neu.edu.cn}
\and
\IEEEauthorblockN{4\textsuperscript{th} Tayaba Yeasmin}
\IEEEauthorblockA{\textit{School of Electrical Engineering and Computer
		Science} \\
	\textit{Washington State University}\\
	Pullman WA, USA\\
tayabayeasmin8228@gmail.com}
}

\maketitle

\begin{abstract}
  In this paper, we study scale-free weak synchronization for
  multi-agent systems (MAS). In other words, we design a protocol for
  the agents without using any knowledge about the network. We do not
  even require knowledge about the connectivity of the network. Each
  protocol contains an adaptive parameter to tune the protocol
  automatically to the demands of the network.
\end{abstract}


\section{Introduction}
Multi-agent systems have been extensively studied over the past 20
years. Initiated by early work such as
\cite{ren-atkins,saber-murray2}, although the roots can be found in
much earlier work \cite{wu-chua2}, it has become an active research
area. But the realization that control systems often consist of many
components with limited or restricted communication between them was
already known much longer and studied in the area of decentralized
control, see e.g. \cite{siljak,corfmat-morse2}. Applications are for
instance systems with many generators connected through a grid or
traffic applications such as platoons of cars. The fallacy of early
decentralized control is that it often created a specific agent which
has a kind of supervisory role while other agents ensure communication
to and from this supervisory agent. This approach turned out to be
highly sensitive to failures in the network.  Multi-agent systems
created a different type of structure in these networks where all
agents basically have a similar role towards achieving synchronization
in the network. However, early work still heavily relied on knowledge
of the network.

Later it was established that the protocols designed for a multi-agent
systems would work for any network structure satisfying some
underlying assumptions such as lower or upper bounds on the spectrum
of the Laplacian matrix associated to the graph describing the network
structure.  In recent years, adaptive protocols were developed to
achieve synchronization problem of MAS, see
\cite{li-wen-duan-ren-tac-2015,lv-li-ren-duan-chen-auto-2016,liu-zhang-saberi-stoorvogel-auto,%
	lv-li-scis-2021}. These protocols get rid of all assumptions on the
network by using time-varying parameters in the protocol, related to
these bounds on the eigenvalues of the Laplacian. However, it still
requires that the network is strongly connected or has a direct
spanning tree. This actually still inherently has some of the
difficulties presented before. How can we check if this connectivity
is present in the network? Secondly, what happens in case of a fault
in the network that makes the network fail this assumption.

In the basic setup of a multi-agent system, the signals exchanged over
the network converge to zero whenever the network synchronizes. So the
fact that the network communication dies out over time is a weaker
condition than output synchronization. We will refer to this weaker
condition as network stability in this paper and a protocol achieving
network stability achieves a form of weak synchronization as
clarified in this paper.

It turns out that synchronization implies network stability and hence
weak synchronization. But, more importantly, if the network has a
directed spanning tree then the converse implication is true: weak
synchronization implies classical synchronization.

We can therefore design adaptive nonlinear protocol which achieve weak
state synchronization for any network without making any kind of
assumptions. If the network happens to have a directed spanning tree
then we obtain classical synchronization. However, if this is not the
case then we describe in detail in this paper what kind of
synchronization properties are preserved in the system. For
applications this kind of weak synchronization yields what one would
hope for. If the cars in a platoon lost connectivity between two
subgroups because their distance has become too large the protocols
will still achieve synchronization in both of these groups. If in a
power system the connectivity between two subgroups is lost, each of
these groups will internally achieve synchronization but, obviously,
no global synchronization will be achieved. However, in general the
decomposition of the network in a case of a fault might be more
involved and these more general cases are also studied in the paper.

\section{Communication network and graph}

To describe the information flow among the agents we associate
a {weighted graph} $\mathcal{G}$ to the communication network. The
weighted graph $\mathcal{G}$ is defined by a triple $(\mathcal{V},
\mathcal{E}, \mathcal{A})$ where $\mathcal{V}=\{1,\ldots, N\}$ is a
node set, $\mathcal{E}$ is a set of pairs of nodes indicating
connections among nodes, and $\mathcal{A}=[a_{ij}]\in
\mathbb{R}^{N\times N}$ is the weighted adjacency matrix with non
negative elements $a_{ij}$. Each pair in $\mathcal{E}$ is called an
{edge}, where $a_{ij}>0$ denotes an edge $(j,i)\in \mathcal{E}$ from
node $j$ to node $i$ with weight $a_{ij}$. Moreover, $a_{ij}=0$ if
there is no edge from node $j$ to node $i$. We assume there are no
self-loops, i.e.\ we have $a_{ii}=0$. A {path} from node $i_1$ to
$i_k$ is a sequence of nodes $\{i_1,\ldots, i_k\}$ such that $(i_j,
i_{j+1})\in \mathcal{E}$ for $j=1,\ldots, k-1$. A {directed tree} is a
subgraph (subset of nodes and edges) in which every node has exactly
one parent node except for one node, called the {root}, which has no
parent node. A {directed spanning tree} is a subgraph which is a
directed tree containing all the nodes of the original graph. If a
directed spanning tree exists, the root of this spanning tree has a
directed path to every other node in the network \cite{royle-godsil}.

For a weighted graph $\mathcal{G}$, the matrix $L=[\ell_{ij}]$
with
\[ \ell_{ij}=
\begin{system}{cl} \sum_{k=1}^{N} a_{ik}, & i=j,\\ -a_{ij}, &
	i\neq j,
\end{system}
\]
is called the {Laplacian matrix} associated with the graph
$\mathcal{G}$. The Laplacian matrix $L$ has all its eigenvalues in the
closed right half plane and at least one eigenvalue at zero associated
with right eigenvector $\textbf{1}$ \cite{royle-godsil}. The zero
eigenvalues of Laplacian matrix is always semi simple, i.e. its
algebraic and geometric multiplicities coincides. Moreover, the graph
contains a directed spanning tree if and only if the Laplacian matrix
$L$ has a single eigenvalue at the origin and all other eigenvalues
are located in the open right-half complex plane \cite{ren-book}.

\begin{figure}[ht] \includegraphics[width=8cm]{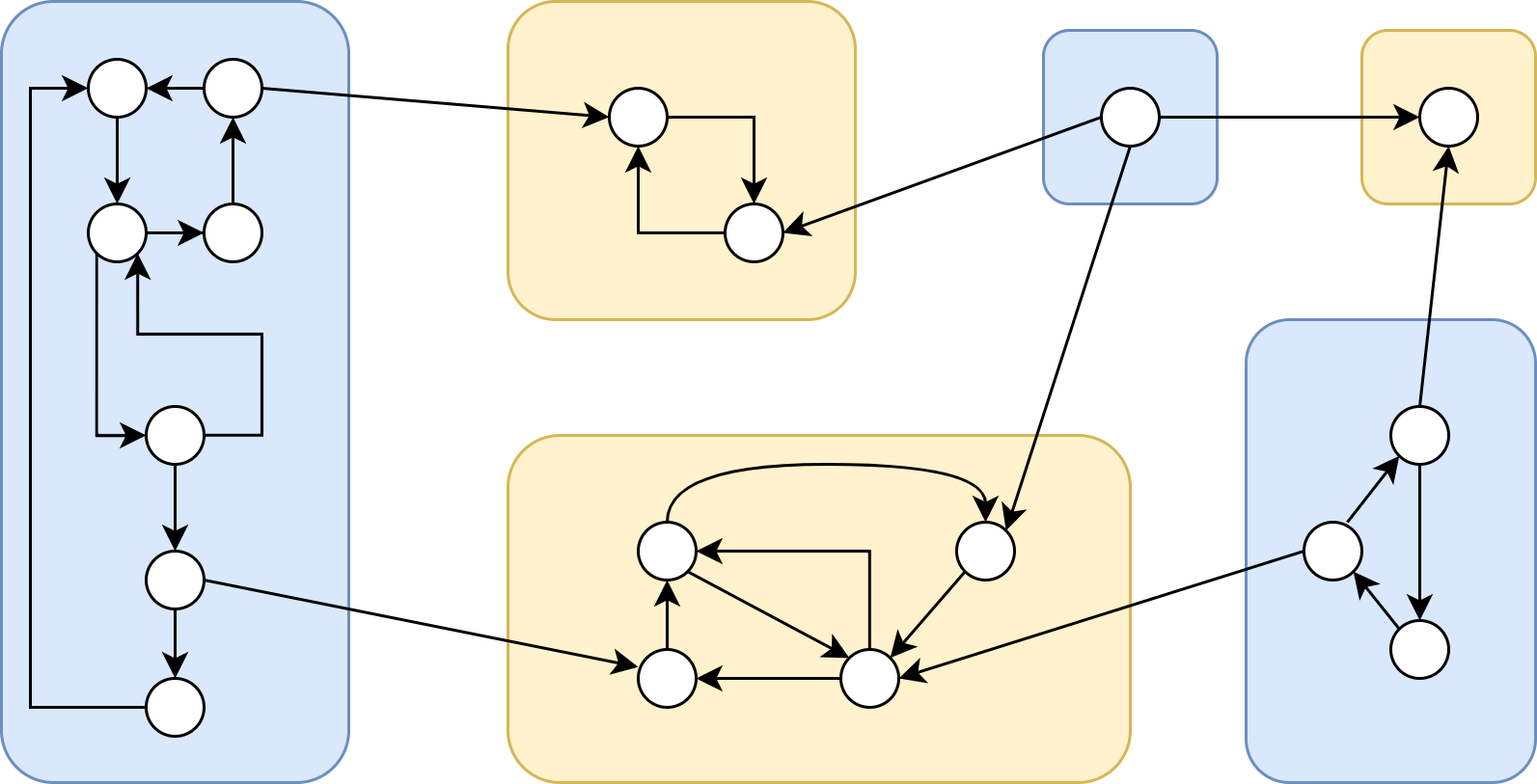}
	\centering
	\caption{A directed communication network and its
		bicomponents.}\label{f1}
\end{figure}

\begin{figure}[ht] \includegraphics[width=8cm]{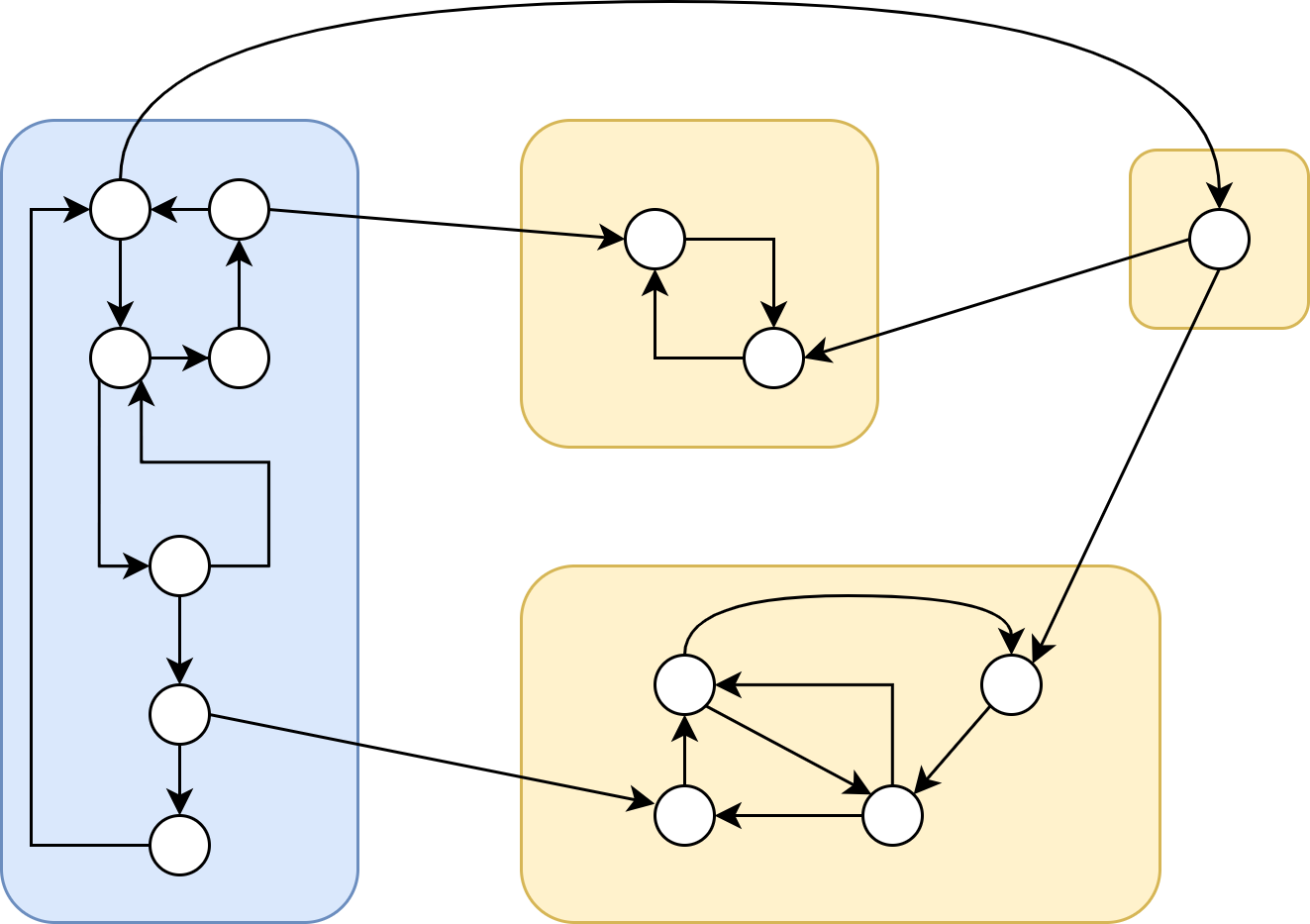}
	\centering
	\caption{A directed communication network with a
		spanning tree and its bicomponents.}\label{f2}
\end{figure}

A directed communication network is said to be strongly connected if
it contains a directed path from every node to every other node in the
graph. For a given graph $\mathcal{G}$ every maximal (by inclusion)
strongly connected subgraph is called a bicomponent of the graph. A
bicomponent without any incoming edges is called a basic
bicomponent. Every graph has at least one basic bicomponent. A network
has one unique basic bicomponent if and only if the network contains
a directed spanning tree. In general, every node in a network can be
reached by at least one basic bicomponent, see \cite[page
7]{stanoev-smilkov-2013}. In Fig. \ref{f1} a directed communication
network with its bicomponents is shown. The network in this figure
contains 6 bicomponents, 3 basic bicomponents (the blue ones) and 3
non-basic bicomponents (the yellow ones). In Fig. \ref{f2} a directed communication
network with its bicomponents is shown. The network in this figure
contains 4 bicomponents but only one basic bicomponent (the blue one).

In the absence of a directed spanning tree, the Laplacian matrix of
the graph has an eigenvalue at the origin with a multiplicity $k$
larger than $1$. This implies that it is a $k$-reducible matrix and
the graph has $k$ basic bicomponents.  The book \cite[Definition
2.19]{wu-book} shows that, after a suitable permutation of the nodes,
a Laplacian matrix with $k$ basic bicomponents can be written in the
following form:
\begin{equation}\label{Lstruc}
  L=\begin{pmatrix} L_0 & L_{01}
    & \cdots & \cdots & L_{0k} \\ 0 & L_1 & 0 & \cdots & 0 \\ \vdots & \ddots &
    \ddots & \ddots & \vdots \\ \vdots & & \ddots & L_{k-1} & 0 \\ 0 &
    \cdots & \cdots & 0 & L_k
  \end{pmatrix}
\end{equation}
where $L_1,\ldots, L_k$ are the Laplacian
matrices associated to the $k$ basic bicomponents in our
network. These matrices have a simple eigenvalue in $0$ because they
are associated with a strongly connected component. On the other hand,
$L_0$ contains all non-basic bicomponents and is a grounded Laplacian
with all eigenvalues in the open right-half plane. After all, if $L_0$
would be singular then the network would have an additional basic
bicomponent.

\section{Weak synchronization of MAS}

In this section, we introduce the concept of weak synchronization for
homogeneous MAS. Consider $N$ homogeneous agents
\begin{equation}\label{system}
  \begin{system*}{cl}
    \dot{x}_i &= Ax_i +B u_i, 
  \end{system*}
\end{equation}
where $x_i\in\mathbb{R}^{n}$ and $u_i\in\mathbb{R}^{m}$ are the state and input of agent $i$ for
$i=1,\ldots, N$ with $(A,B)$ stabilizable and $(C,A)$ detectable.

The communication network provides agent $i$ with the following
information which is a linear combination of its own output relative
to that of other agents
\begin{equation}\label{zeta1}
  \zeta_i=\sum_{j=1}^{N}a_{ij}(x_i-x_j),
\end{equation}
where $a_{ij}\geq 0$ and $a_{ii}=0$. The communication topology of the
network can be described by a weighted and directed graph
$\mathcal{G}$ with nodes corresponding to the agents in the network
and the weight of edges given by coefficient $a_{ij}$. In terms of the
coefficients of the associated Laplacian matrix $L$, $\zeta_i$ can be
rewritten as
\begin{equation}\label{zeta}
  \zeta_i= \sum_{j=1}^{N}\ell_{ij}x_j.
\end{equation}
We denote by $\mathbb{G}^N$ the set of all graphs with $N$ nodes.  

Our protocols are of the form:
\begin{equation}\label{protocoln}
  \begin{system*}{ccl}
    \dot{\xi}_i &=& f_i(\xi_i, \zeta_i) \\
    u_i &=& g_i(\xi_i, \zeta_i)
  \end{system*}
\end{equation}

In the following, we introduce the concept of network stability which
is an intrinsically different concept compared to state synchronization.

\begin{definition}[\textbf{Network stability}] \label{wea0}
  Consider a multi-agent network described by \eqref{system},
  \eqref{zeta}. If the protocol \eqref{protocoln} is such that
  \[
    \zeta_i(t)=\sum_{j=1}^{N}a_{ij}(x_i-x_j)\to 0
  \]
  as $t\to \infty$, for any $i \in \{1,\ldots, N\}$ and for all
  possible initial conditions, then the protocol achieves network
  stability/
\end{definition}

\begin{definition}\label{wea}
  Consider an MAS described by \eqref{system} and \eqref{zeta1}, with
  protocols of the form \eqref{protocoln}. The network achieves state
  synchronization if the states of the respective agents satisfy:
  \begin{equation}\label{syncod}
    x_i(t)-x_j(t) \rightarrow 0
  \end{equation}
  as $t\rightarrow \infty$ for any $i,j\in \{1,\ldots, N\}$ and for
  all possible initial conditions.
\end{definition}

Next, we present two lemmas to explain the difference between these two
kind of synchronization

\begin{lemma}\label{xxxxx}
  Consider an MAS described by \eqref{system} and \eqref{zeta1}.  with
  protocols of the form \eqref{protocol}. In that case output
  synchronization implies network stability.
\end{lemma}

\begin{lemma}\label{xxxxx2}
  Consider an MAS described by \eqref{system} and
  \eqref{zeta1}. Assume the protocols \eqref{protocoln} achieves
  network stability. In that case the network achieves weak
  synchronization in the sense that:
  \begin{itemize}
  \item If the network contains a directed spanning tree then we
    always achieve output synchronization.
  \item If the network does not contain a directed spanning tree then
    we only achieve output synchronization in the trivial case where
    all agents are asymptotically stable and synchronization is
    obvious.
  \item Assume the network does not have a directed spanning tree
    which implies that the graph has $k>1$ basic bicomponents.
    \begin{itemize}[label=\ding{212}]
    \item Within basic bicomponent $i$ we have output synchronization
      in the sense that there exists a signal $x^i_s$ such that
      $x_j(t)-x^i_s(t)\rightarrow 0$ provided agent $j$ is part of
      basic bicomponent $i$.
    \item An agent $j$ which is not part of any of the basic
      bicomponents synchronizes to a trajectory $y_{j,s}$,
      \begin{equation}\label{ys}
        x_{j,s}=\sum_{i=1}^k\, \beta_{j,i} x^i_s 
      \end{equation}
      where the coefficients $\beta_{j,i}$ are nonnegative, satisfy:
      \begin{equation}\label{betasum}
        1=\sum_{i=1}^k\, \beta_{j,i} 
      \end{equation}
      and only depend on the parameters of the network and do not depend
      on any of the initial conditions.
    \end{itemize}
  \end{itemize}
\end{lemma}

The proofs for Lemmas 1 and 2 are some minor adaptation of the proof of the
same results in \cite[Lemmas 1 and 2]{stoorvogel-saberi-liu-weak} to include
nonlinear protocols.

\section{Problem formulation}

In this section, we consider the homogeneous agents of the form
\eqref{system} with network communication given by \eqref{zeta1} or, equivalently,
\eqref{zeta}.

We make the following assumption.

\begin{assumption}\label{ass}\mbox{}
  The pair $(A,B)$ is stabilizable.
\end{assumption}

Next, we define the problem addressed in this paper:

\begin{problem}\label{prob_x}
  Consider a MAS \eqref{system} with associated network communication
  \eqref{zeta}. The \textbf{scale-free weak state synchronization} is
  to find, if possible, a fully distributed nonlinear protocol
  \eqref{protocoln} using only knowledge of agent models, i.e.,
  $(A, B)$, such that the MAS with the above protocol achieves
  scale-free weak state synchronization, i.e.\ for any graph
  $\mathscr{G}\in\mathbb{G}^N$ with any size of the network $N$, the
  MAS achieves network stability as defined in Definition
  \ref{wea0}.
\end{problem}

\section{Protocol design for homogeneous MAS}

We design an adaptive protocol to achieve the
objectives of Problem \ref{prob_x} for this model \eqref{system}
through two steps as following.

\begin{tcolorbox}[colback=white]
  \textbf{Step 1. Find matrix} $\mathbf{P}$\textbf{:}\\
  Under the assumption that $(A,B)$ is stabilizable, there exists a
  matrix $P>0$ satisfying the following algebraic Riccati equation
  \begin{equation}\label{eq-Riccati}
    A\T P + PA -PBB\T P + I =0.
  \end{equation}
  \textbf{Step 2. Obtain protocol}\textbf{:}\newline We design the
  following adaptive protocol
  \begin{equation}\label{protocol}
    \begin{system*}{ccl}
      \dot{\rho}_i &=& \zeta_i\T PBB\T P \zeta_i \\
      u_i &=& -\rho_i B\T P \zeta_i.
    \end{system*}
  \end{equation}
\end{tcolorbox}

We have the following theorem.

\begin{theorem}\label{theorem}
  Consider a homogeneous MAS \eqref{system} with associated network
  communication \eqref{zeta} satisfying Assumption \ref{ass}. The
  \textbf{scale-free weak state synchronization for homogeneous MAS}
  as stated in problem \ref{prob_x} is solvable. In particular,
  protocol \eqref{protocol} achieves network stability for an
  arbitrary number of agents $N$ and for any graph
  $\mathscr{G}\in\mathbb{G}^N$.
\end{theorem}

Before we can proof the above theorem we derive two crucial lemmas.

\begin{lemma}\label{lem1}
	Consider a number of agents $N$ and a graph
	$\mathscr{G}\in\mathbb{G}^N$. Consider MAS \eqref{system} with
	associated network communication \eqref{zeta}. Assume
	Assumption \ref{ass} is satisfied.  If all $\rho_i$ remain bounded
	then we have:
	\begin{equation}\label{probdef}
		\zeta_i(t) \rightarrow 0
	\end{equation}
	$t\rightarrow \infty$ for all $i=1,\ldots, N$.
\end{lemma}

\begin{proof}[Proof of Lemma \ref{lem1}]
	We define
	\[
	x=\begin{pmatrix} x_1 \\ \vdots \\ x_N \end{pmatrix},
	\]
	and
	\[
	\rho=\begin{pmatrix}
		\rho_1 & 0      & \cdots & 0 \\
		0      & \rho_2 & \ddots & \vdots \\
		\vdots & \ddots & \ddots & 0 \\
		0      & \cdots & 0      & \rho_N
	\end{pmatrix}
	\]
	We obtain
	\begin{equation}\label{syscomp}
		\dot{x}=(I\otimes A)x-(\rho L\otimes BB\T P)x,
	\end{equation}
	and we define:
	\[
	\zeta_i = (L_i \otimes I)x,
	\]
	where $L_i$ is the $i$'th row of $L$ for $i=1,\ldots, N$.  We obtain
	\begin{equation}\label{barxt3}
		\zeta = (L \otimes I) x. 
	\end{equation}
	where
	\[
	\zeta=\begin{pmatrix} \zeta_1 \\ \vdots \\ \zeta_N \end{pmatrix}.
	\]
	By using \eqref{syscomp}, we obtain
	\begin{equation}\label{barxt4}
		\dot{\zeta} = (I\otimes A)\zeta
		-[L\rho \otimes BB\T
		P]\zeta. 
	\end{equation}
	For any $i\in \{1,\ldots, N\}$, from \eqref{barxt4} we have that
	\begin{equation*}
		\dot{\zeta}_i = A\zeta_i
		-[L_i\rho \otimes BB\T P]\zeta. 
	\end{equation*} 
	Define
	\[
	V_i = \zeta_i\T P \zeta_i,
	\]
	then we obtain
	\begin{equation}\label{barxt7}
		\dot{V}_i = -\zeta_i\T \zeta_i + \zeta_i\T PBB\T P \zeta_i
		-2\zeta_i\T \left[ L_{i}\rho \otimes PBB\T P
		\right] \zeta. 
	\end{equation}
	Since we know
	\[
	\dot{\rho}_i = r_i\T r_i
	\]
	where
	\[
	r_i = B\T P \zeta_i.
	\]
	we find that $\rho_i$ bounded implies that $r_i\in L_2$.
	Moreover, we have:
	\begin{equation}\label{si}
		s_i =-2\left[ L_i \rho \otimes
		B\T P \right] \zeta = -2 \sum_{j=1}^N \ell_{ij} \rho_{j}r_j
	\end{equation}
	which yields $s_i\in L_2$.  We obtain from \eqref{barxt7} that
	\begin{equation}\label{barxt8}
		\dot{V}_i \leq -\eta V_i + r_i\T r_i + r_i\T s_i \leq -\eta V_i +
		2r_i\T r_i + 2s_i\T s_i 
	\end{equation}
	where $\eta= \| P \|^{-1}$. Since $r_i$ and $s_i$ are both
	in $L_2$ this implies $V_i(t) \rightarrow 0$ which yields that
	$\zeta_i(t)\rightarrow 0$. Since this is true for any
	$i \in \{1,\ldots,N\}$ we find that we achieve weak state
	synchronization.
\end{proof}   

In the first Lemma we showed that if all the adaptive parameters
remain bounded then we obtain our desired result. The next lemma
establishes that the adaptive parameters are actually bounded and we
can use Lemma \ref{lem1} to establish synchronization.

\begin{lemma}\label{lem2a}
	Consider MAS \eqref{system} with associated network
	communication \eqref{zeta} and the protocol \eqref{protocol}.
	Assume Assumption \ref{ass} is satisfied. Additionally, assume that
	either the $\rho_i$ associated to agents belonging to the basic
	bicomponents are bounded or the graph is strongly connected.In that
	case, all $\rho_i$ remain bounded.
\end{lemma}

\begin{proof}  
	Using the notation of Lemma \ref{lem1} we obtain \eqref{syscomp}.
	We first consider the case that some but not all of the $\rho_i$ are
	unbounded. Without loss of generality, we renumber the agents such
	that $\rho_i$ is unbounded for $i\leq k$ while $\rho_i$ is bounded
	for $i>k$ with $k<N$. We have
	\begin{multline*}
		L = \begin{pmatrix}
			L_{11} & L_{12} \\ L_{21} & L_{22} 
		\end{pmatrix},\quad
		x^k =\begin{pmatrix} x_1 \\ \vdots \\ x_k \end{pmatrix}, \quad
		x^k_c =\begin{pmatrix} x_{k+1} \\ \vdots \\
			x_{N} \end{pmatrix},\\
		\zeta^k =\begin{pmatrix} \zeta_{1} \\ \vdots \\
			\zeta_{k} \end{pmatrix},\quad
		\zeta^k_c =\begin{pmatrix} \zeta_{k+1} \\ \vdots \\
			\zeta_{N} \end{pmatrix},
	\end{multline*}
	with $L_{11}\in \R^{k\times k}$. If all the agents associated to
	basic bicomponents have a bounded $\rho_i$ this implies that agents
	associated to $i=1,\ldots,k$ are not associated to basic
	bicomponents which implies that $L_{11}$ is invertible. On the other
	hand, if the network is strongly connected we always have $L_{11}$
	is invertible since $k<N$. We have
	\begin{equation}\label{K1K2}
		\| v^k_c \|_2 < K_1.
	\end{equation}
	for suitably chosen $K_1$ where
	\[
	v^k_c = (I \otimes B\T P) \zeta^k_c
	\]
	which follows from the fact that when $\rho_i$ is bounded for
	$i=k+1,\ldots, N$ then $v_i\in L_2$ (note that
	$\dot{\rho}_i = v_i\T v_i$). We define
	\[
	\hat{x}^k= x^k+(L_{11}^{-1}L_{12}\otimes I) x^k_c.
	\] 
	Using \eqref{syscomp} we then obtain
	\begin{equation}\label{barxt2xxx}
		\dot{\hat{x}}^k = (I\otimes A) \hat{x}^k
		-[\rho^{k} L_{11} \otimes BB\T
		P]\hat{x}^k  -\left[L_{11}^{-1}L_{12}\rho^k_c
		\otimes B\right]v^k_c
	\end{equation}
	where 
	\[
	\rho^k=\begin{pmatrix}
		\rho_1 & 0      & \cdots & 0 \\
		0      & \rho_2 & \ddots & \vdots \\
		\vdots & \ddots & \ddots & 0 \\
		0      & \cdots & 0      & \rho_k
	\end{pmatrix},
	\rho^k_c=\begin{pmatrix}
		\rho_{k+1} & 0      & \cdots & 0 \\
		0      & \rho_{k+2} & \ddots & \vdots \\
		\vdots & \ddots & \ddots & 0 \\
		0      & \cdots & 0      & \rho_N
	\end{pmatrix}.
	\]
	Define
	\[
	\hat{v}^k =-(L_{11}^{-1}L_{12}\rho^k_c \otimes I)v^k_c , 
	\]
	then \eqref{K1K2} in combination with the boundedness of
	$\rho^k_c$ implies that there exists $K_2$ such that
	\begin{equation}\label{K3K4}
		\| \hat{v}^k \|_2 < K_2.
	\end{equation}
	We obtain
	\begin{equation}\label{barxt2xxxy}
		\dot{\hat{x}}^k = (I\otimes A)\hat{x}^k
		-[\rho^k L_{11} \otimes BB\T
		P]\hat{x}^k+[I \otimes B]\hat{v}^k
	\end{equation}
	and we define:
	\begin{equation}\label{Vj}
		V_k= (\hat{x}^k)\T (\rho^{-k} H^k \otimes P) \hat{x}^k,
	\end{equation}
	with $\rho^{-k}=(\rho^k)^{-1}$ while:
	\begin{equation}\label{HN}
		H^k=\begin{pmatrix}
			\alpha_1 & 0      & \cdots & 0 \\
			0      & \alpha_2 & \ddots & \vdots \\
			\vdots & \ddots & \ddots & 0 \\
			0      & \cdots & 0      & \alpha_k
		\end{pmatrix}.
	\end{equation}
	By \cite[Theorem 4.25]{qu-book-2009} we can choose
	$\alpha_1,\ldots,\alpha_k>0$ such that $H^k L_{11} +L_{11}\T H^k > 0$. It is
	easily seen that this implies that there exists a $\gamma$ such
	that:
	\begin{equation}\label{HkL11}
		H^kL_{11}+L_{11}\T H^k > 3\gamma L_{11}\T L_{11},
	\end{equation}
	We get from \eqref{barxt2xxxy} that
	\begin{multline*}
		\dot{V}_k \leq (\hat{x}^k)\T \left[ \rho^{-k} H^k \otimes (-I+PBB\T P) \right]
		\hat{x}^k\\ -(\hat{x}^k)\T \left[ (H^kL_{11}+L_{11}\T H^k)
		\otimes PBB\T P \right] \hat{x}^k \\
		+ 2(\hat{x}^k)\T \left[ \rho^{-k} H^k \otimes PB\right]\hat{v}^k.
	\end{multline*}
	where we used that $V_k$ is decreasing in $\rho_i$ for
	$i=1,\ldots, k$. The above yields for $t>T$ that
	\begin{multline*}
		\dot{V}_k \leq -V_k 
		- 2\gamma (\hat{x}^k)\T \left[ L_{11}\T L_{11} \otimes PBB\T P \right] \hat{x}^k \\
		+ 2(\hat{x}^k)\T \left[\rho^{-k} H^k \otimes PB\right]\hat{v}^k,
	\end{multline*}
	provided $T$ is such that
	\[
	\rho^{-2k} (H^k)^2  < \gamma L_{11}\T L_{11},
	\]
	for $t>T$ which is possible since we have
	$\rho_i\rightarrow \infty$ for $i=1,\ldots, k$. We define
	\[
	\check{v}^k = \left[ L_{11}^{-1}
	\rho^{-k}H^k \otimes I\right]\hat{v}^k. 
	\]
	We get
	\begin{equation}\label{lastone2xx}
		\dot{V}_k \leq - V_k 
		- \gamma (\hat{x}^k)\T \left[ L_{11}\T L_{11} \otimes PBB\T P \right] \hat{x}^k 
		+(\check{v}^k)\T\check{v}^k,
	\end{equation}
	for $t>T$. Moreover, 
	\begin{equation}\label{rfgt}
		(\hat{x}^k)\T \left[ L_{11}\T L_{11} \otimes PBB\T
		P \right] \hat{x}^k \geq  \sum_{i=1}^k \dot{\rho}_i,
	\end{equation}
	since $(L_{11}\otimes I)\hat{x}^k=\zeta^k$. Hence \eqref{lastone2xx} implies
	\begin{equation}\label{ttg1}
		\dot{V}_k \leq -V_k -\gamma
		\sum_{i=1}^k \dot{\rho}_{i} + (\check{v}^k)\T\check{v}^k  
	\end{equation} 
	Since the $\rho_i$ are unbounded while $\check{v}^k\in L_2$ this
	yields a contradiction since $V_k\geq 0$.
	
	Next we consider the case that all $\rho_i$ are unbounded. In this
	case, we assumed the graph is strongly connected and hence by Lemma
	\ref{2.8} presented in the appendix there exists
	$\alpha_1,\ldots,\alpha_N>0$ such that \eqref{Hlyap} is satisfied
	with $H^N$ given by \eqref{HN} for $k=N$.  We define:
	\begin{equation}\label{VN}
		V_N= x\T \left[ Q_{\rho} \otimes P 
		\right] x,
	\end{equation}
	where
	\begin{equation}\label{Qrho}
		Q_{\rho} = \rho^{-N} \left( H^N\rho^N - \mu_N
		\textbf{h}_N\textbf{h}_N\T \right) \rho^{-N}
	\end{equation}
	with $\rho^{-N}=(\rho^N)^{-1}$, while
	\[
	\mu_N=\frac{1}{\sum_{i=1}^N \alpha_i\rho_i^{-1}},\qquad
	\textbf{h}_N = \begin{pmatrix} \alpha_1 \\ \vdots \\ \alpha_N \end{pmatrix}.
	\]
	From Lemma \ref{2.9} in the appendix we know that $Q_{\rho}$ is
	decreasing in $\rho_i$ for $i=1,\ldots N$.  Note that
	$Q_\rho \rho L=H^N L$.  We get from \eqref{syscomp} that
	\begin{multline}\label{eqref}
		\dot{V}_N \leq x\T \left[ Q_\rho \otimes (-I+PBB\T P) \right] x\\
		-x\T \left[ (H^NL+L\T H^N) \otimes PBB\T P \right] x 
	\end{multline}
	It is easily verified that $Q_{\rho}\textbf{1}=0$. Moreover
	$\Ker L = \Span \{ \textbf{1} \}$ since the network is strongly
	connected. Therefore
	\[
	\ker Q_{\rho} \subset \Ker L\T L
	\]
	Together with the fact that $\rho_j\rightarrow \infty$ for
	$j=1,\ldots, N$ and therefore $Q_{\rho}\rightarrow 0$ this implies
	that there exists $T$ such that
	\begin{equation}\label{Qrhobound}
		Q_{\rho} < \gamma L\T L,
	\end{equation}
	is satisfied for $t>T$. 
	
	The above together with \eqref{Hlyap} yields for $t>T$ that
	\begin{equation*}
		\dot{V}_N \leq -\eta V_N
		- 2\gamma x\T \left[ L\T L \otimes PBB\T P \right] x 
	\end{equation*}
	where $\eta=\| P\|^{-1}$. Note that
	\[
	x\T \left[ L\T L \otimes PBB\T
	P \right] x  = \sum_{i=1}^N \dot{\rho}_i,
	\]
	The above implies
	\begin{equation}\label{ttg1N}
		\dot{V}_N \leq -\eta V_N -\gamma \sum_{i=1}^N \dot{\rho}_{i} 
	\end{equation} 
	This again yields a contradiction since the $\rho_i$ are unbounded
	while $V_N \geq 0$.
\end{proof}

\begin{proof}[Proof of Theorem \ref{theorem}]
	The Laplacian matrix of the system in general has the form
	\eqref{Lstruc}. We note that if we look at the dynamics of the
	agents belonging to one of the basic bicomponents then these
	dynamics are not influenced by the other agents and hence can be
	analyzed independent of the rest of network. The network within one
	of the basic bicomponents is strongly connected and we can apply
	Lemmas \ref{lem2a} to guarantee that the $\rho_i$ associated to a
	basic bicomponents are all bounded.
	
	Next, we look at the full network again. We have already established
	that the $\rho_i$ associated to all basic bicomponents are all
	bounded. Then we can again apply Lemmas \ref{lem2a} to conclude that
	the other $\rho_i$ not associated to basic bicomponents are also
	bounded.
	
	After having established that all the $\rho_i$ are bounded, we can
	then apply Lemma \ref{lem1} to conclude that we achieve scale-free
	weak state synchronization.
\end{proof}

\begin{figure}[ht]
	\includegraphics[width=5cm]{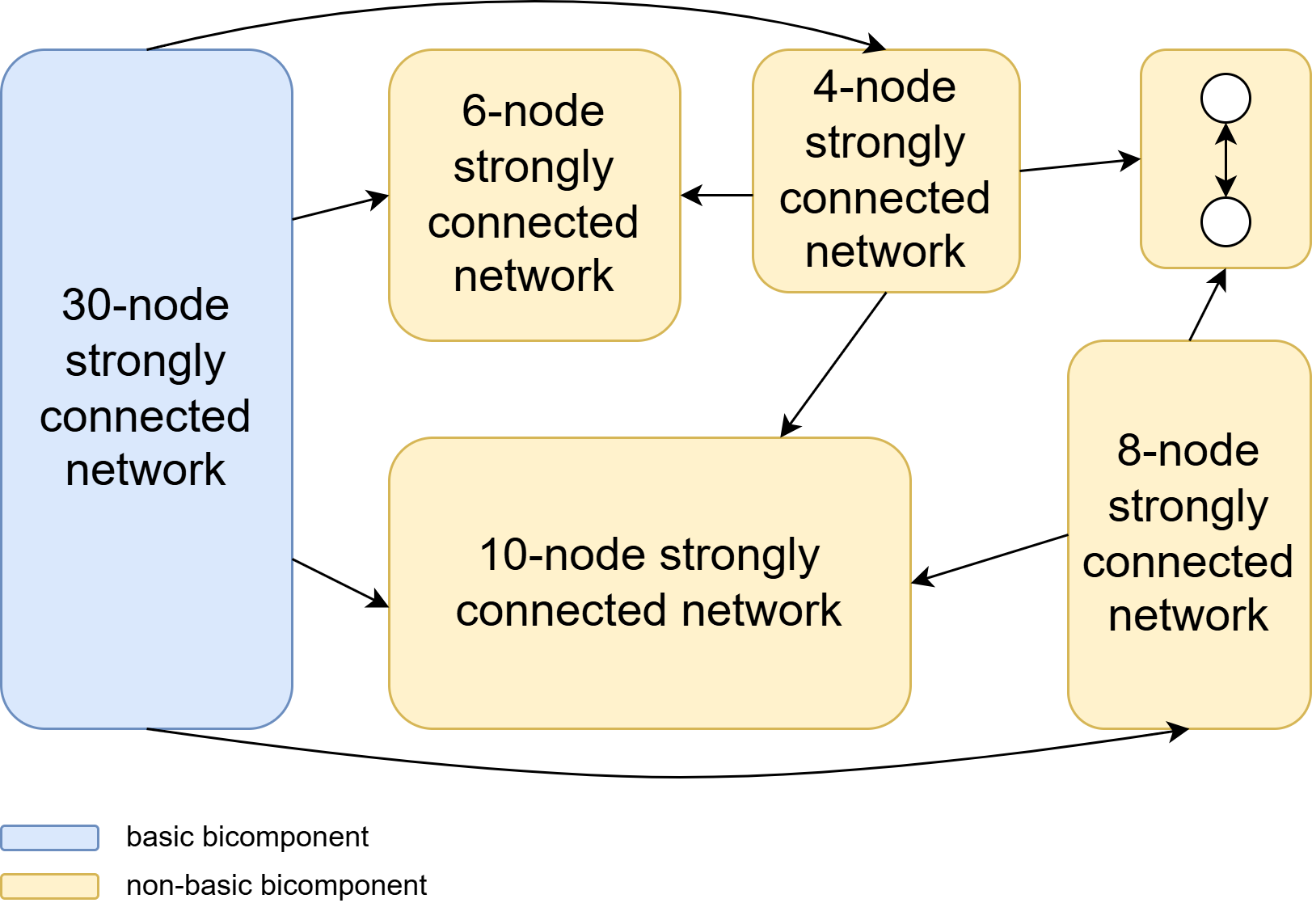}
	\centering
	\caption{The 60-nodes communication network with spanning tree.}\label{f5}
\end{figure}
\begin{figure}[ht]
	\includegraphics[width=8cm]{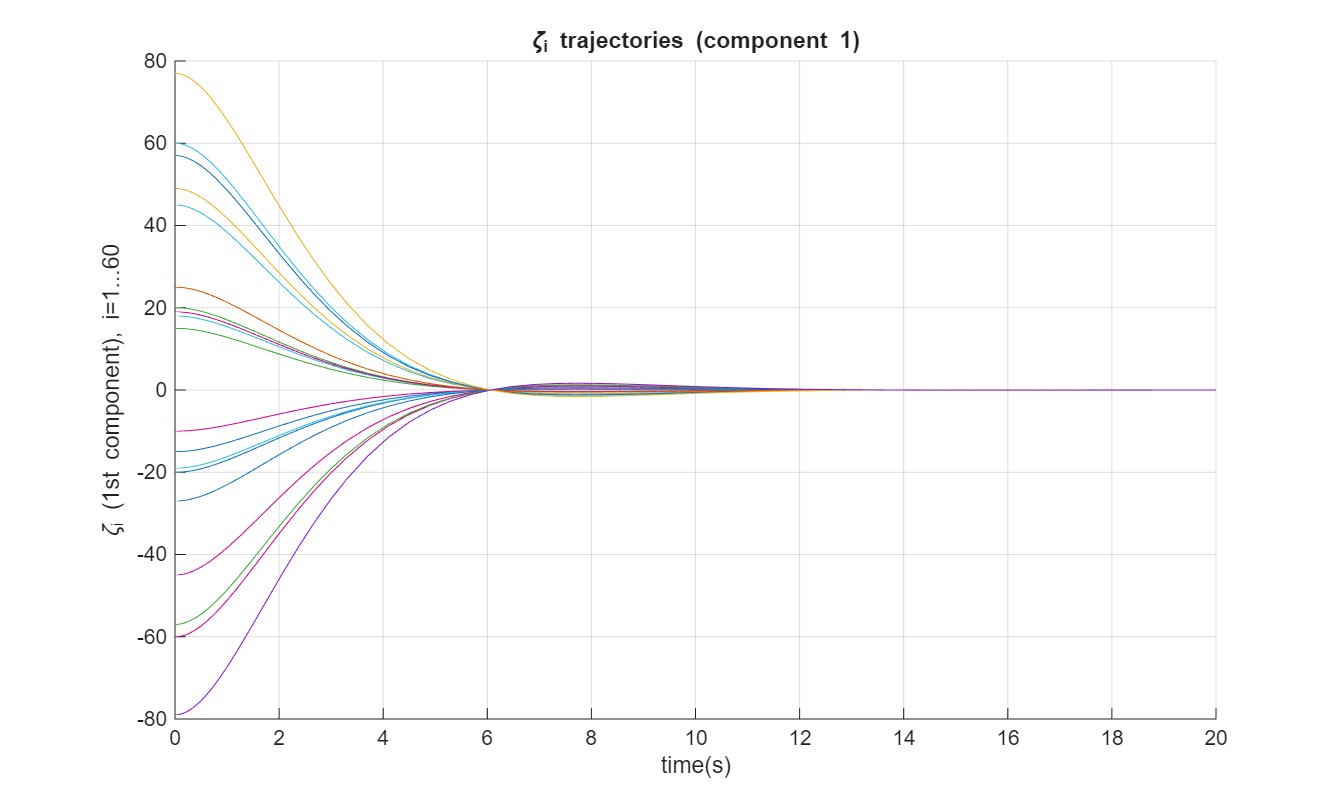}
	\centering
	\caption{The trajectory of $\zeta_i$ for graph containing spanning
		tree.}\label{zeta60tree}
\end{figure}
\begin{figure}[ht]
	\includegraphics[width=8cm]{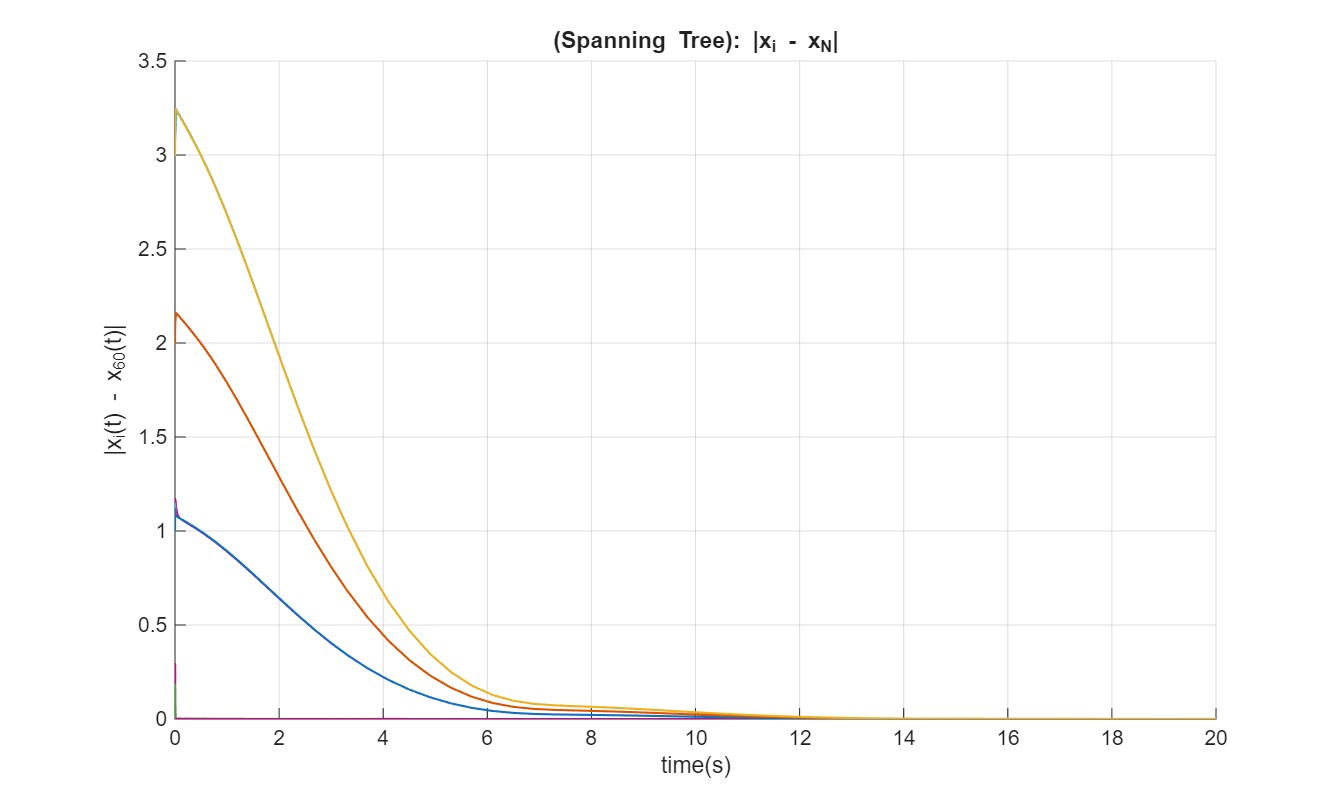}
	\centering
	\caption{The trajectory of agents' states.}\label{statesyn60tree}
\end{figure}
\begin{figure}[ht]
	\includegraphics[width=8cm]{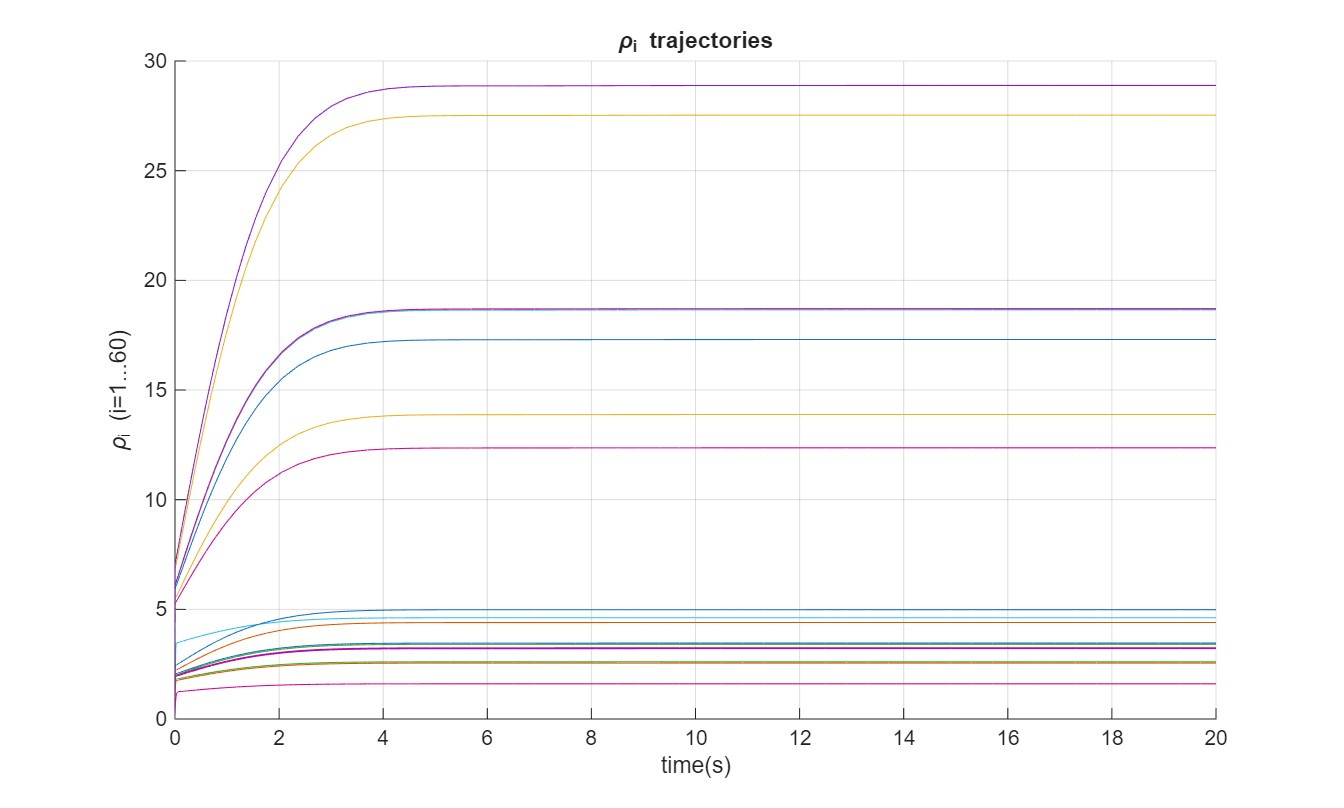}
	\centering
	\caption{The trajectory of $\rho_i$.}\label{rho60tree}
\end{figure}
\begin{figure}[ht]
	\includegraphics[width=5cm]{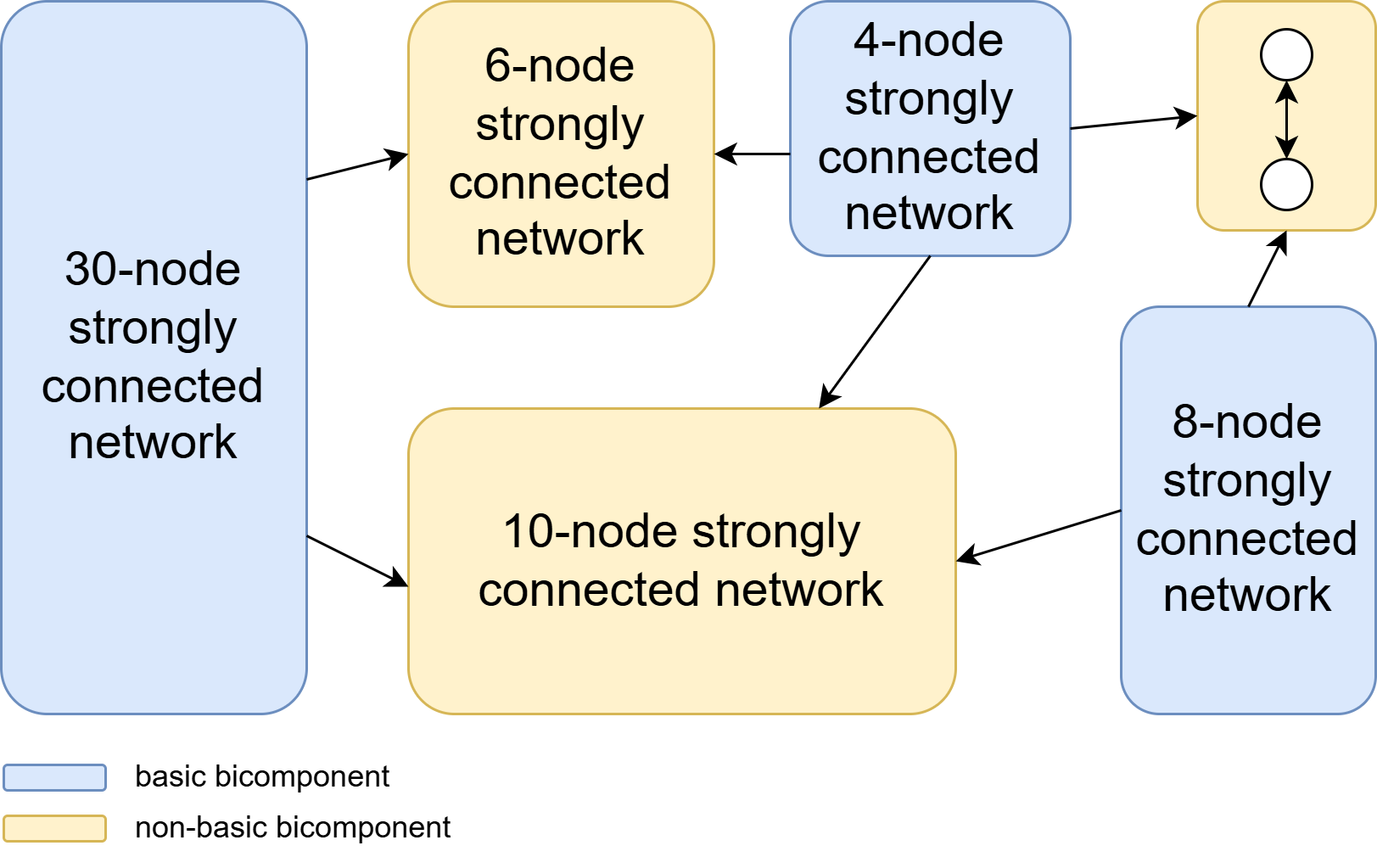}
	\centering
	\caption{The communication network without spanning tree. The links
		are broken due to faults. }\label{f4}
\end{figure}
\begin{figure}[ht]
	\includegraphics[width=8cm]{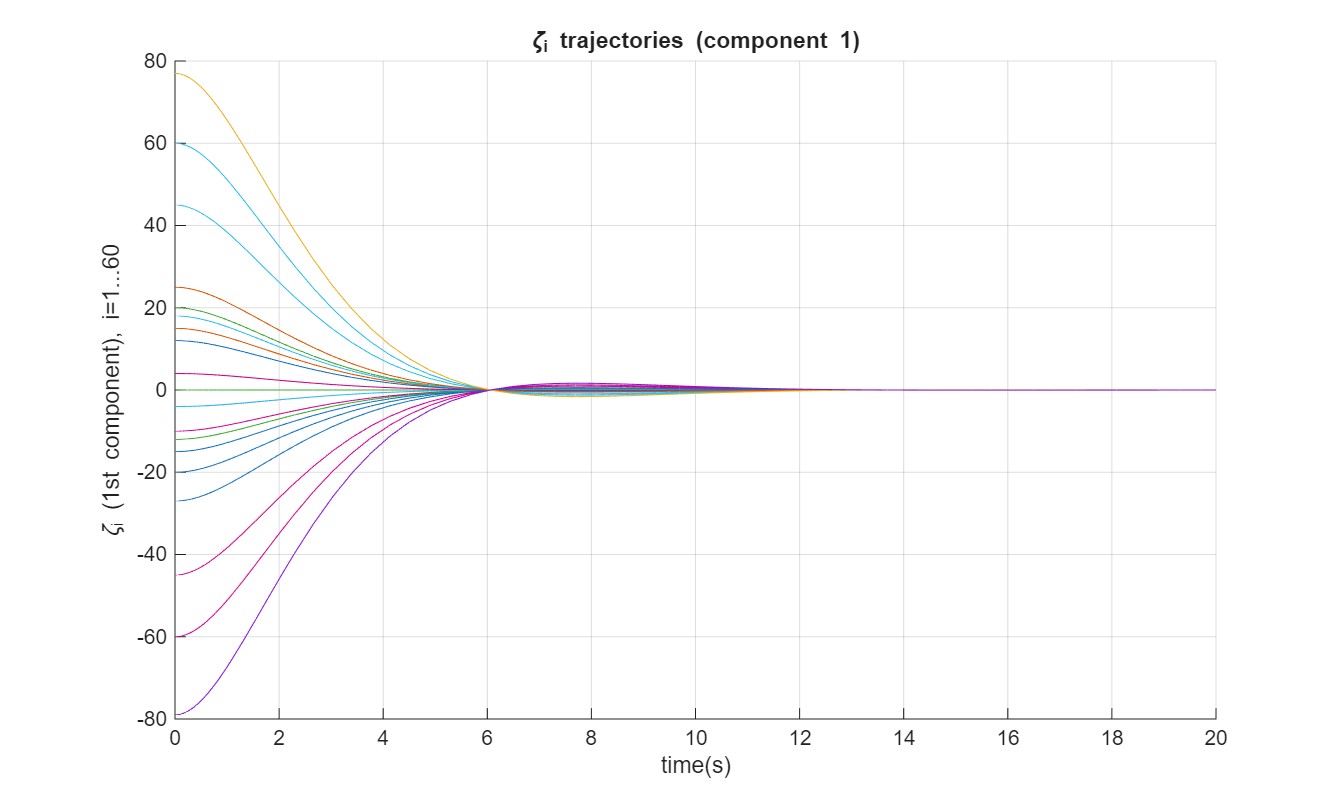}
	\centering
	\caption{The trajectory of $\zeta_i$ for graph without directed
		spanning tree.}\label{zeta60n} 
\end{figure}
\begin{figure}[ht]
	\includegraphics[width=8cm]{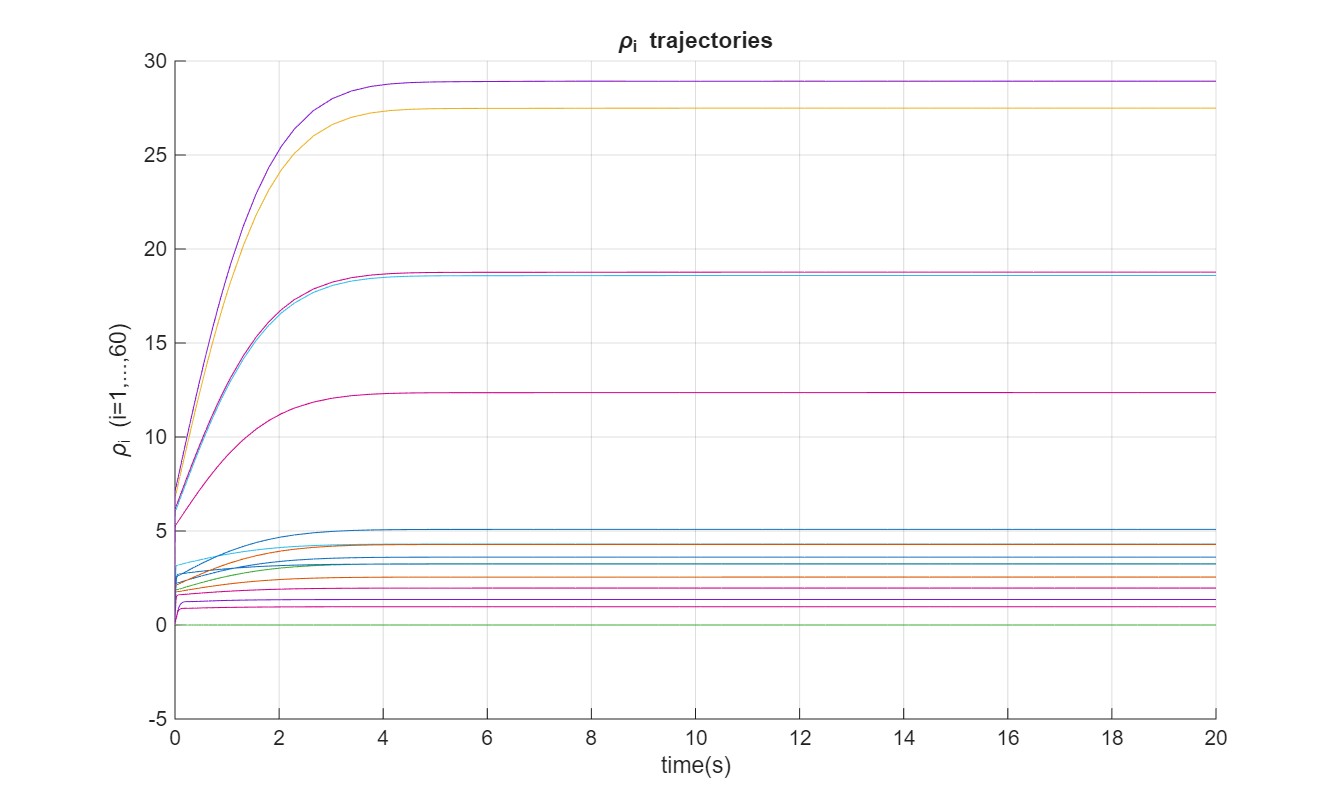}
	\centering
	\caption{The trajectory of $\rho_i$ for graph without directed
		spanning tree.}\label{rho60n} 
\end{figure}
\begin{figure}[ht]
	\includegraphics[width=8cm]{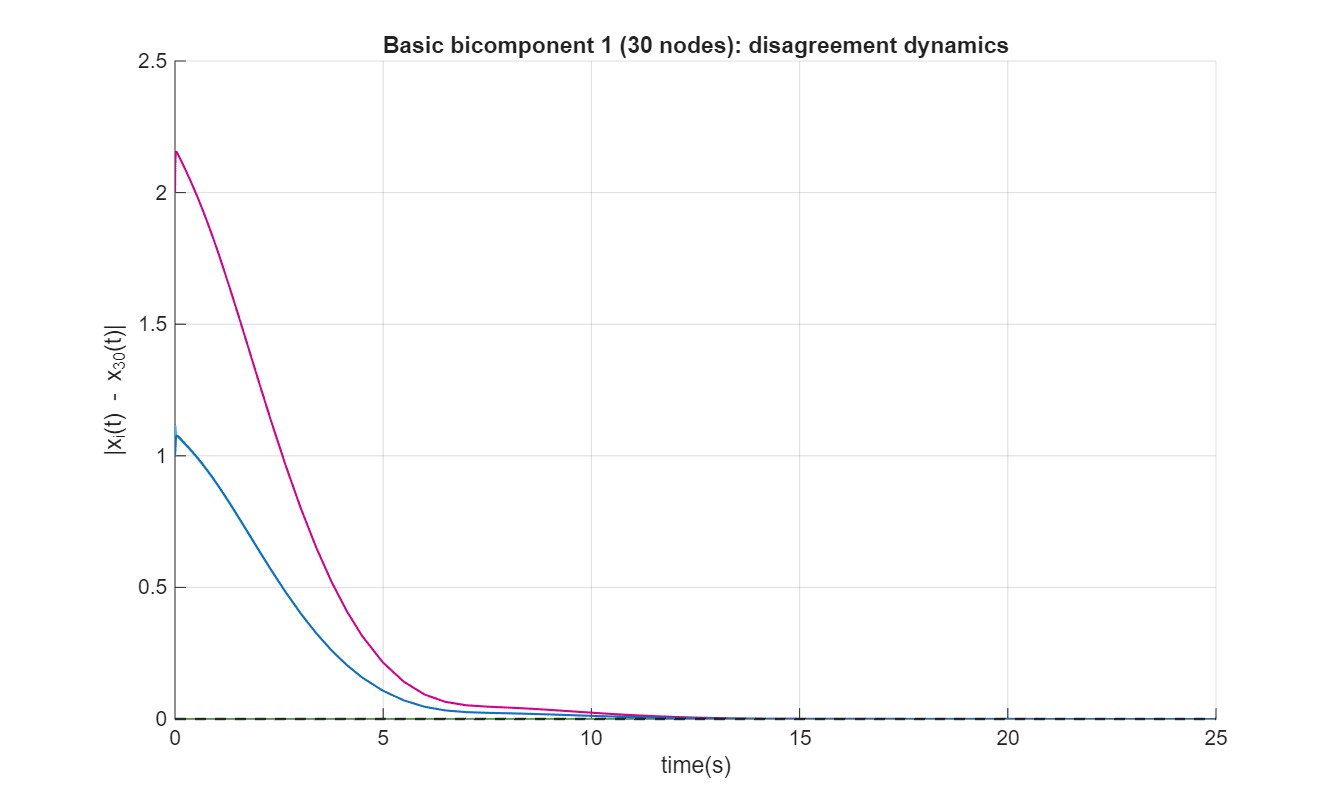}
	\centering
	\caption{Basic bicomponent 1 (30 nodes): the disagreement dynamic among
		the agents.}\label{ct30co}
\end{figure}
\begin{figure}[ht]
	\includegraphics[width=8cm]{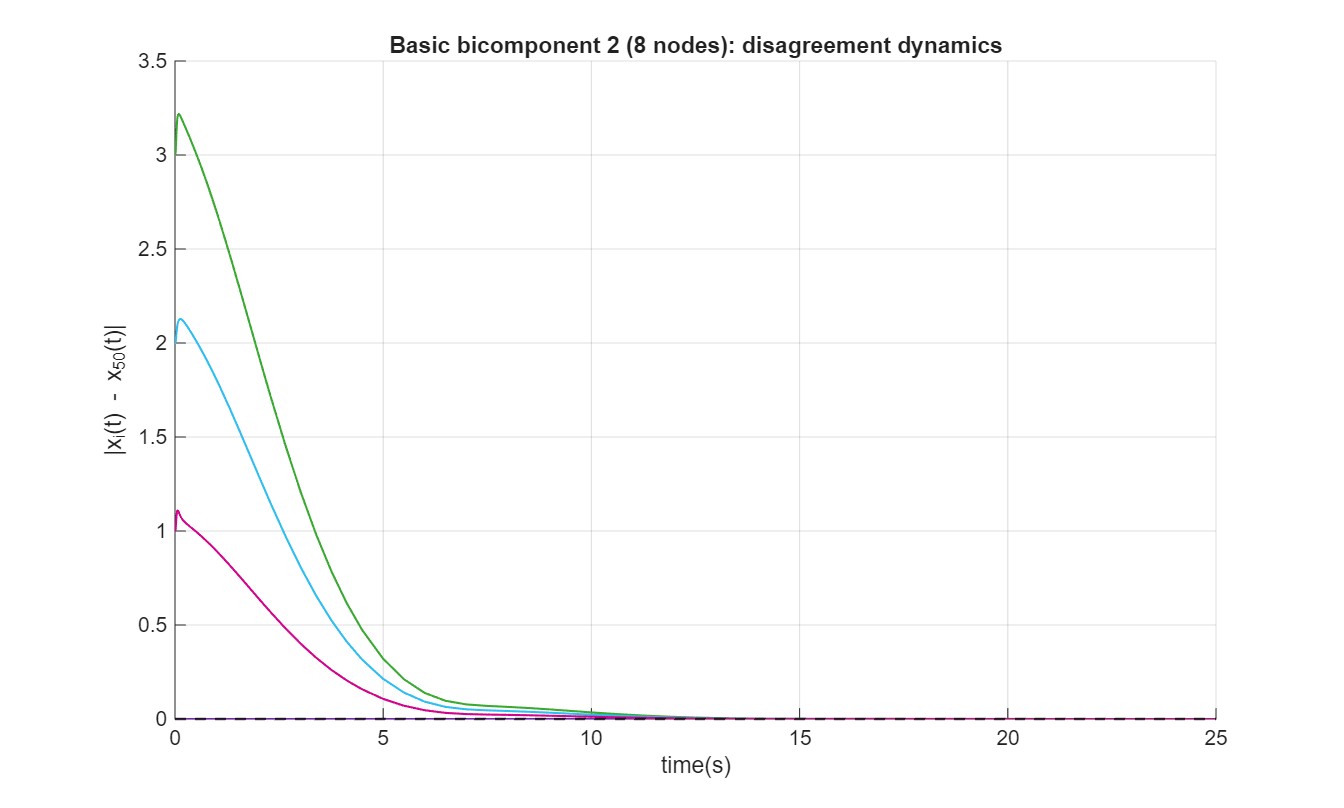} \centering
	\caption{Basic bicomponent 2 (8 nodes): the disagreement dynamic among
		the agents.}\label{ct8co}
\end{figure}
\begin{figure}[ht]
	\includegraphics[width=8cm]{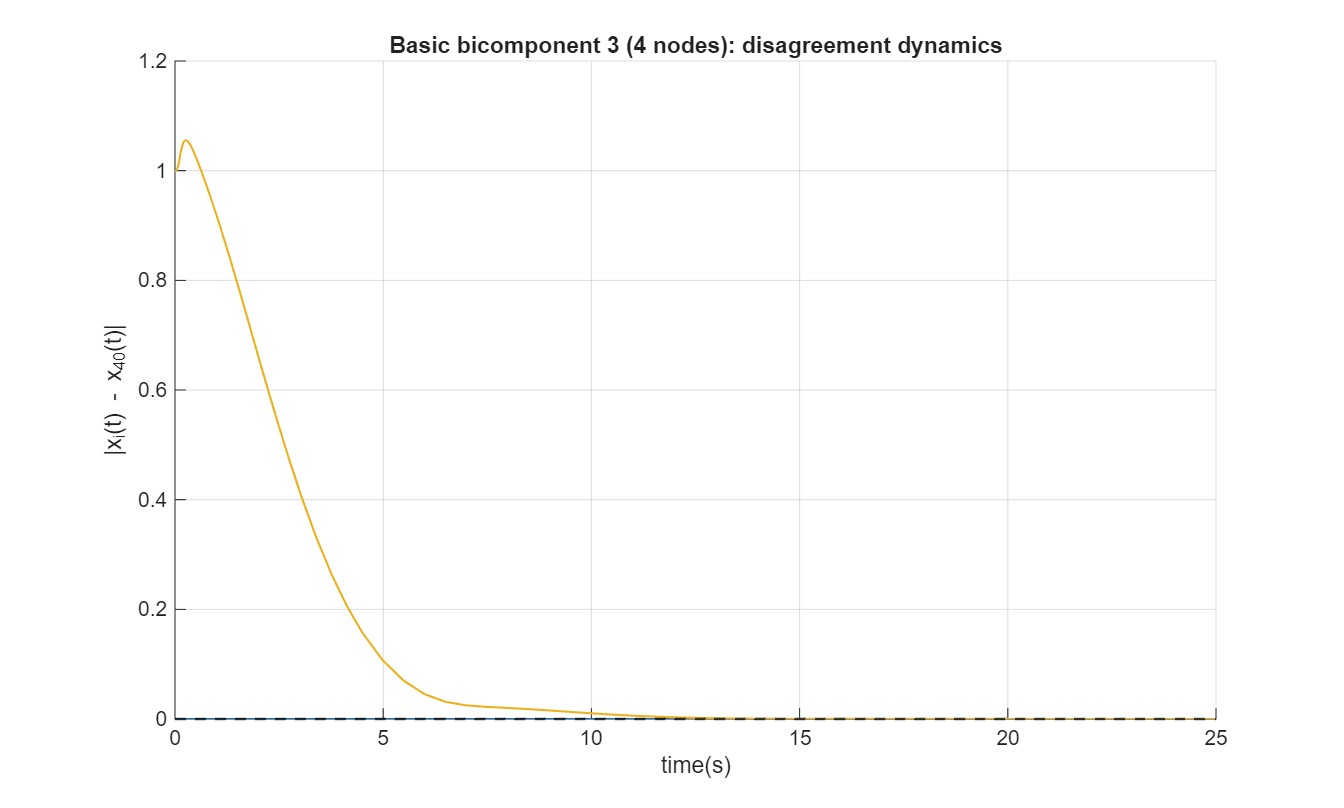}
	\centering
	\caption{Basic bicomponent 3 (4 nodes): the disagreement dynamic among
		the agents.}\label{ct4co}
\end{figure}
\begin{figure}[ht]
	\includegraphics[width=5cm]{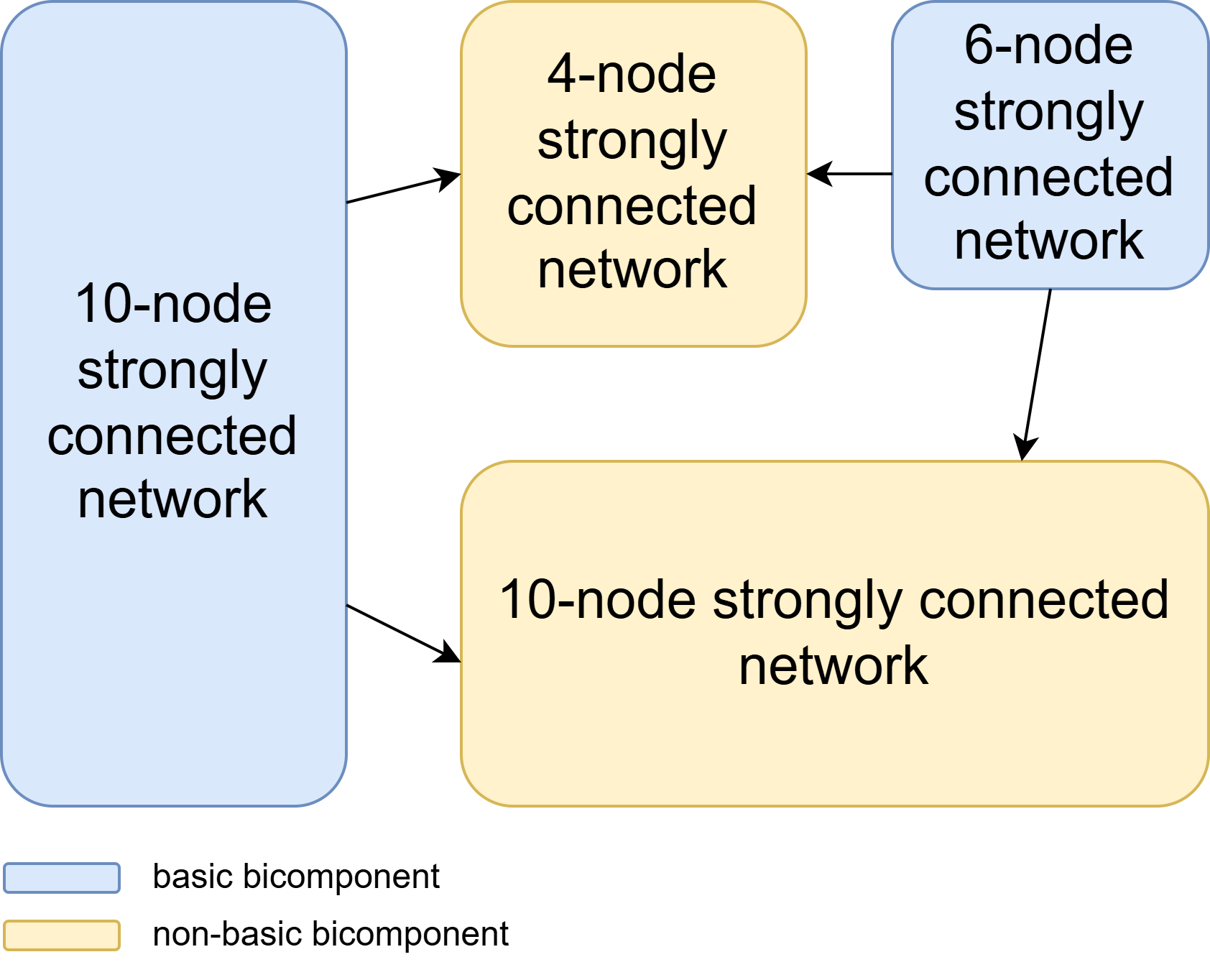}
	\centering
	\caption{The 30-node communication network without spanning tree. The
		links are broken due to faults. }\label{f8}  
\end{figure}
\begin{figure}[ht]
	\includegraphics[width=8cm]{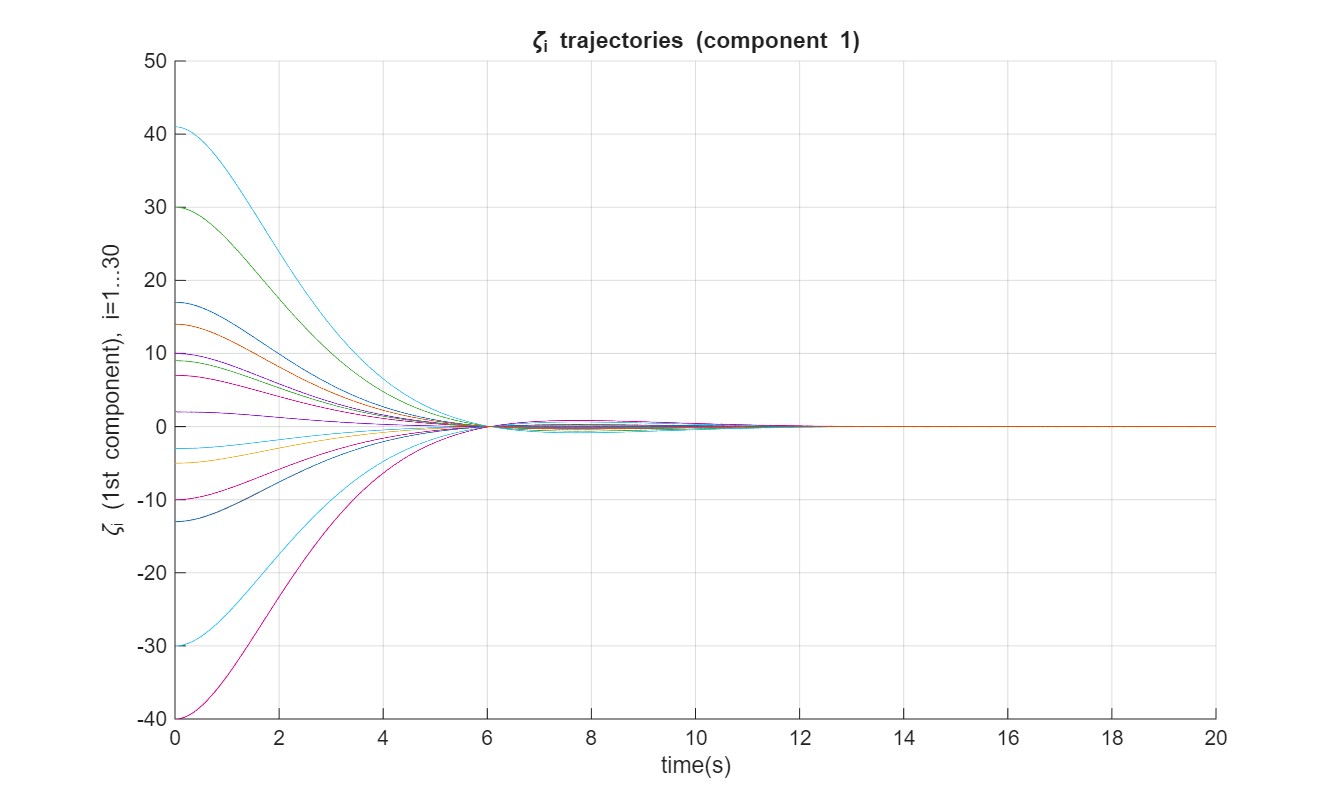}
	\centering
	\caption{The trajectory of $\zeta_i$ for 30-node graph without
		directed spanning tree.}\label{zeta30} 
\end{figure}
\begin{figure}[ht]
	\includegraphics[width=8cm]{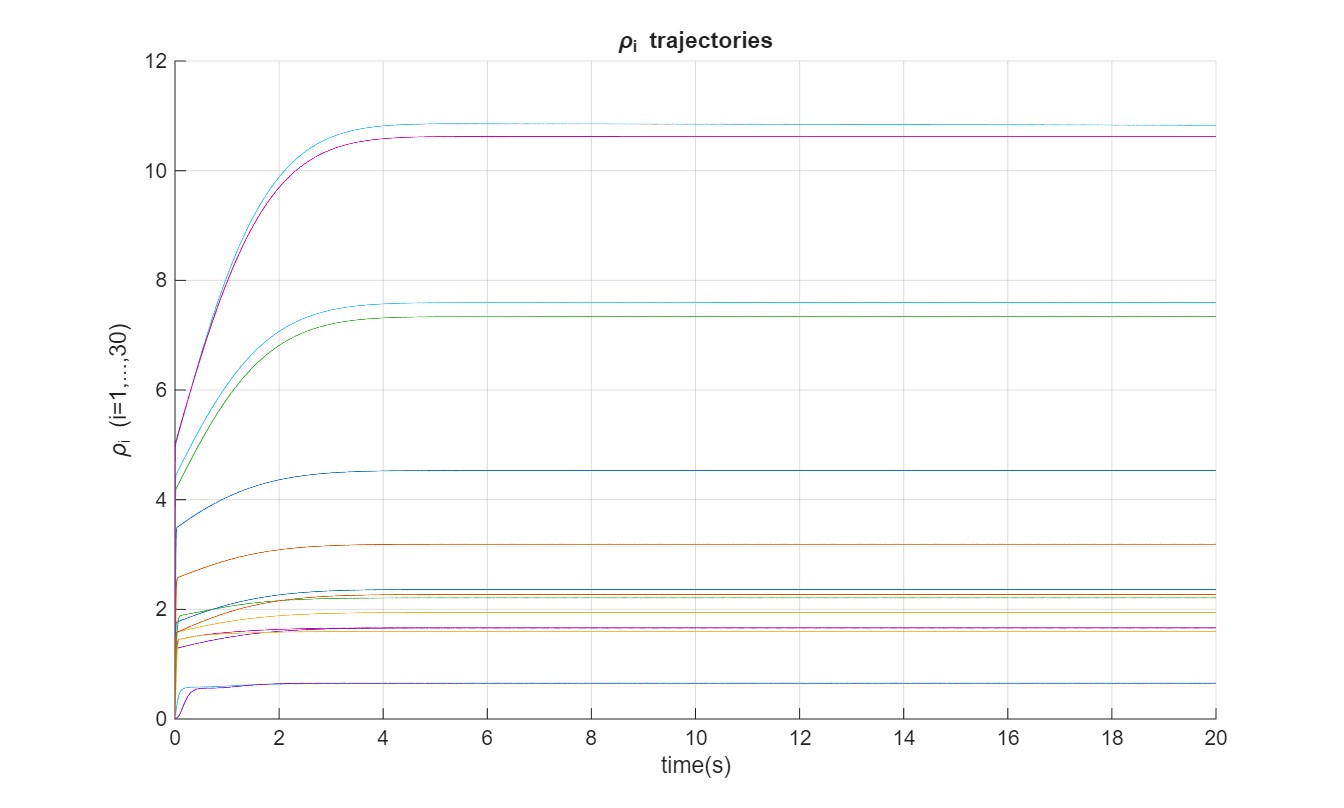}
	\centering
	\caption{The trajectory of $\rho_i$ for 30-node graph without
		directed spanning tree.}\label{rho30} 
\end{figure}
\begin{figure}[ht]
	\includegraphics[width=8cm]{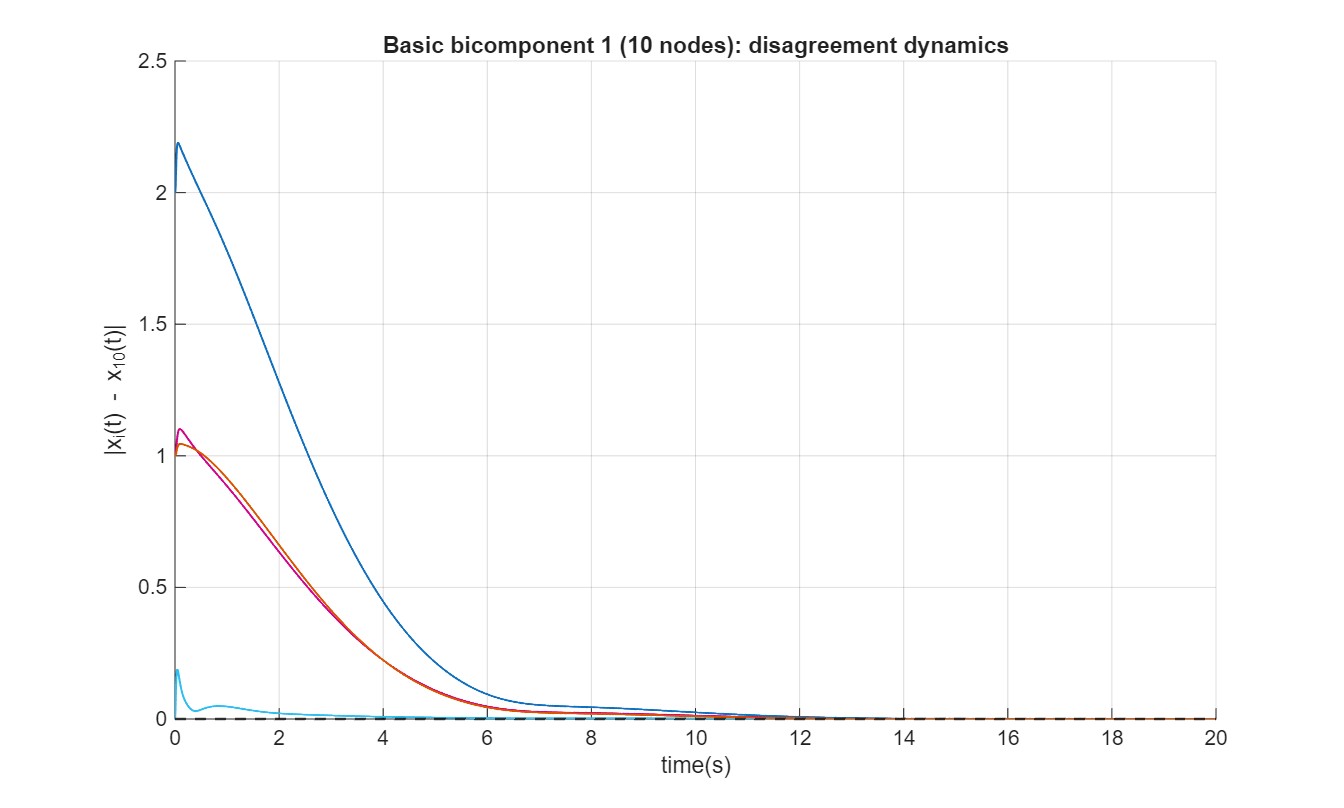}
	\centering
	\caption{Basic bicomponent 1 (10 nodes) in 30-node graph: the
		disagreement dynamic among the agents.}\label{30-10}
\end{figure}
\begin{figure}[ht]
	\includegraphics[width=8cm]{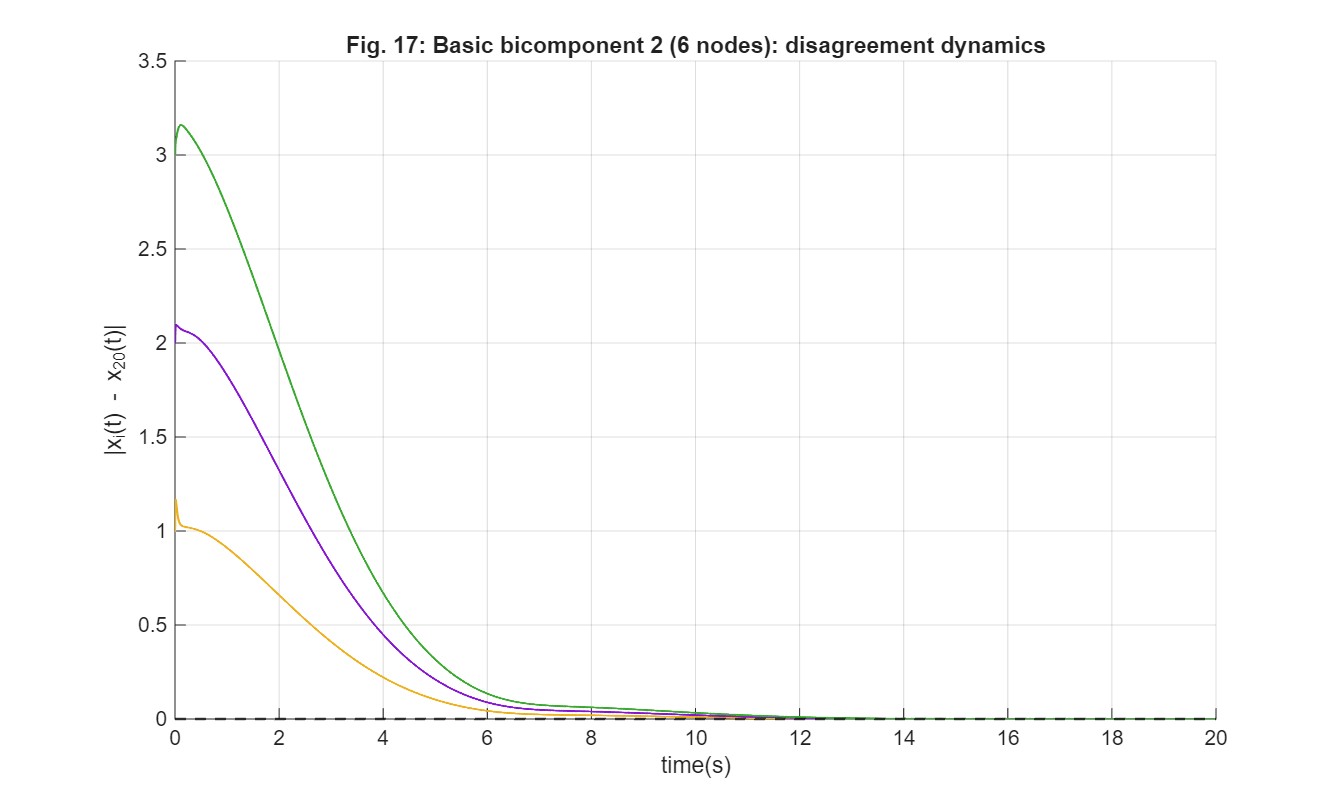} 
	\centering
	\caption{Basic bicomponent 2 (6 nodes) in 30-node graph: the
		disagreement dynamic among the agents.}\label{30-6}
\end{figure}

\section{Numerical examples}

In this section, we consider agent models of the form \eqref{system} with the following parameters,
\begin{equation*}
	A=\begin{pmatrix}
		0&1&0\\0&0&1\\0&0&0
	\end{pmatrix},B=\begin{pmatrix}
		0\\0\\1
	\end{pmatrix}.
\end{equation*}

By using the protocol in \eqref{eq-Riccati} and \eqref{protocol}, we
can obtain the following adaptive protocol:
\begin{tcolorbox}[colback=white]
	\begin{equation}\label{protocol-s}
		\begin{system}{l}
			\dot{\rho}_i = \zeta_i\T {\setlength{\arraycolsep}{2mm}\begin{pmatrix}
					1&2.41&2.41\\
					2.41&5.82&5.82\\
					2.41&5.82&5.82
			\end{pmatrix}} \zeta_i \\
			u_i = -\rho_i {\setlength{\arraycolsep}{2mm}\begin{pmatrix}
					1&2.41&2.41
			\end{pmatrix}} \zeta_i.
		\end{system}
	\end{equation}
	with 
	\[
	P=\begin{pmatrix}
		2.41&2.41&1\\
		2.41&4.82&2.41\\
		1&2.41&2.41
	\end{pmatrix}.
	\]
\end{tcolorbox}

\subsection{Network containing a directed spanning tree}

We consider state synchronization result for the 60-node homogeneous
network shown in figure \ref{f5}, which contains a directed spanning
tree. By using the adaptive protocol \eqref{protocol-s}, we obtain
$\zeta_i\to 0$ and $\dot{\rho}_i$ are bounded, which means state synchronization is
achieved when the network contains a directed spanning tree, see
Figs. \ref{zeta60tree} and \ref{rho60tree}.

\subsection{Network without a directed spanning tree}

When some links have faults, the communication network
might lose its directed spanning tree. For example, if two specific
links are broken in the original 60-node network given by Figure
\ref{f5}, then we obtain the network as given in Figure \ref{f4}

It is obvious that there is no spanning tree in Figure \ref{f4}. We
obtain three basic bicomponents (indicated in blue): one containing 30
nodes, one containing 8 nodes and one containing 4 nodes. Meanwhile,
there are three non-basic bicomponents: one containing 10 nodes, one
containing 6 nodes and one containing 10 nodes, which are indicated in
yellow.

To show the proposed protocol can achieve the scale-free weak
synchronization, we will provide 2 cases with 2 different graphs in
this subsection.

\subsubsection{Case I: 60-node graph shown in Fig. \ref{f4}}

By using the adaptive protocol \eqref{protocol-s}, we obtain
$\zeta_i\to 0$ as $t\to \infty$ for the graph shown in Fig. \ref{f4},
which means weak synchronization is achieved in the absence of
connectivity, see Fig. \ref{zeta60n}. And then, the adaptive parameters
$\rho_i$ are also bounded, see Fig. \ref{rho60n}. It implies that the
available network data for each agent goes to zero and the
communication network becomes inactive.

We have seen that for the 60-node network given in Figure \ref{f5}
this adaptive protocol indeed achieves state synchronization. If we apply the
same protocol to the network described by Figure \ref{f4} which does
not contain a directed spanning tree, we again consider the six
bicomponents constituting the network. 
We see that, consistent with
the theory, we get state synchronization within the three basic
bicomponents as illustrated in Figures \ref{ct30co}, \ref{ct8co} and
\ref{ct4co} respectively. Clearly, the disagreement dynamic among
the agents (the errors between the state of agents) goes to zero
within each basic bicomponent.

%
%
%
%
%
%

\subsubsection{Case II: 30-node graph shown in Fig. \ref{f8}}

For the above 30-node graph, we obtain $\zeta_i\to 0$ as $t\to \infty$
for the graph shown in Fig. \ref{f8} by using the adaptive protocol
\eqref{protocol-s}, i.e., weak synchronization is achieved in the
absence of connectivity, see Fig. \ref{zeta30}. And then, the adaptive
parameters $\rho_i$ are also bounded, see Fig. \ref{rho30}. It implies
that the available network data for each agent goes to zero and the
communication network becomes inactive.

Meanwhile, we obtain state synchronization within the three basic
bicomponents as illustrated in Figures \ref{30-10} and \ref{30-6}
respectively. Clearly, the disagreement dynamic among the agents (the
errors between the state of agents) goes to zero within each basic
bicomponent.

From these examples, we found that the protocol \eqref{protocol-s} can
achieve the scale-free weak synchronization, which means that there
exist $\zeta_i\to 0$ as $t\to \infty$ for any graph
$\mathscr{G}\in\mathbb{G}^N$ with arbitrary number of agents $N$. And
the common synchronization can be still achieved in the basic
bicomponents of the graph.

\section{Conclusion}

In this paper we have introduced network stability and weak state
synchronization for MAS and we have seen that this properties can be
achieved via an adaptive nonlinear protocol. If we have a directed
spanning tree then we obtain the classical concept of state
synchronization. If, for instance due to a fault, the network no
longer contains a directed spanning tree, then we still achieve
network stability which implies a weaker form of synchronization, weak
synchronization. Weak synchronizations guarantees a stable response to
these faults: within basic bicomponents we still achieve
synchronization and the states of agents not contained in a basic
bicomponent converge to a convex combination of the asymptotic
behavior achieved in the basic bicomponents.

\bibliographystyle{IEEEtran}
\bibliography{referenc}

\newcounter{equation2}
\setcounter{equation2}{\value{equation}}
\appendix
\setcounter{equation}{\value{equation2}}%
\renewcommand{\theequation}{\arabic{equation}}

\section*{Some useful lemmas}

\begin{lemma}\label{2.8}
	Consider a directed graph with Laplacian $L$ which is strongly
	connected. Then there exists $\alpha_1,\ldots,\alpha_N>0$ such that:
	\begin{equation}\label{Hlyap}
		H^NL +L\T H^N \geq 3\gamma L\T L, 
	\end{equation}
	for some $\gamma >0$ with $H^N$ given by \eqref{HN} for $k=N$.
\end{lemma}

\begin{proof}
	Choose a left eigenvector of the Laplacian $L$ associated with
	eigenvalue $0$:
	\[
	\begin{pmatrix} \alpha_1 & \cdots & \alpha_N \end{pmatrix} L =0
	\]
	Because the network is strongly connected, by \cite[Theorem
	4.31]{qu-book-2009} we can choose $\alpha_1,\ldots,\alpha_N>0$ and
	obtain:
	\begin{equation}\label{app1}
		H^NL +L\T H^N \geq 0
	\end{equation}
	Note that $H^N L$ has the structure of a Laplacian matrix with a
	zero row sum but also a zero column sum. The latter implies that
	$L\T H^N$ also has the structure of a Laplacian matrix. The sum of
	two Laplacian matrices still has the structure of a Laplacian
	matrix. In other words, $H^NL +L\T H^N$ has the structure of a
	Laplacian matrix. Note that this a undirected graph and there is an
	edge between nodes $i$ and $j$ if there is either an edge from node
	$i$ to $j$ or an edge from node $j$ to $i$ in the original graph
	associated with the Laplacian matrix $L$. Since the graph associated
	with the Laplacian matrix $L$ was strongly connected this implies
	that the graph associated with the Laplacian matrix $H^NL +L\T H^N$
	is also strongly connected. But then the rank of the matrix
	$H^NL +L\T H^N$ is equal to $N-1$ and hence we obtain:
	\begin{equation}\label{app2}
		\Ker (H^NL +L\T H^N ) = \Ker L\T L = \Span \{ \textbf{1} \}
	\end{equation}   
	\eqref{app1} and \eqref{app2} together imply that there exists a
	$\gamma$ such that \eqref{Hlyap} is satisfied.
\end{proof}

We also need to prove that the Lyapunov function in our paper is
decreasing in $\rho$ which is established in the following lemma:

\begin{lemma}\label{2.9}
	The quadratic form:
	\[
	V=z\T Q_{\rho} z
	\]
	with $Q_{\rho}$ given by \eqref{Qrho} is decreasing in $\rho_i$ for
	$i=1,\ldots, k$.
\end{lemma}

\begin{proof}
	Note that
	\begin{equation}\label{appstar1}
		Q_{\rho} \textbf{1} =0
	\end{equation}
	since $\textbf{h}_N\T \rho^{-N} \textbf{1} = \mu_N^{-1}$. Define:
	\[
	z=\begin{pmatrix} z_1 \\ \vdots \\ z_N \end{pmatrix},\qquad
	\bar{z} =\begin{pmatrix} z_1-z_N \\ \vdots \\
		z_{k-1}-z_N \end{pmatrix},\qquad
	\textbf{h}_{N-1} =\begin{pmatrix} \alpha_1\\ \vdots \\
		\alpha_{N-1}\end{pmatrix}
	\]
	and we find:
	\[
	V = \bar{z}\T (\rho^{N-1})^{-1} (\rho^{N-1}H^{N-1} - \mu_N
	\textbf{h}_{N-1}\textbf{h}_{N-1}\T ) (\rho^{N-1})^{-1} \bar{z}
	\]
	using \eqref{appstar1}. Some simple algebra establishes that:
	\begin{multline*}
	(\rho^{N-1})^{-1} (\rho^{N-1}H^{N-1} - \mu_N \textbf{h}_{N-1}\textbf{h}_{N-1}\T )
	(\rho^{N-1})^{-1} \\= \left[ \rho^{N-1}(H^{N-1})^{-1} + \alpha_N^{-1}\rho_N
	\textbf{1}_{N-1}\textbf{1}_{N-1}\T \right]^{-1}
	\end{multline*}
	and hence:
	\[
	V= \bar{z}\T \left[ \rho^{N-1}(H^{N-1})^{-1} + \alpha_N^{-1}\rho_N
	\textbf{1}_{N-1}\textbf{1}_{N-1}\T \right]^{-1} \bar{z}
	\]
	which clearly establishes that $V$ is decreasing in  $\rho_i$ for
	$i=1,\ldots, k$.
\end{proof}

\end{document}